\documentstyle[preprint,aps]{revtex}
\def\L{${\cal L}$ }
\def\be{\begin{equation}}
\def\ee{\end{equation}}
\def\bea{\begin{eqnarray}}
\def\eea{\end{eqnarray}}
\def\ba{\begin{array}}
\def\ea{\end{array}}
\def\a{\alpha}
\def\b{\beta}
\def\c{\gamma}
\def\d{\delta}

\def\0{$\Gamma_0$}

\def\s{\sigma}

\begin{document}
\draft
\title{ New  sum rule identities
and duality relation for  the Potts $n$-point correlation function}
 \author{F. Y. Wu and H. Y. Huang}
\address{Department of Physics,
	 Northeastern University, Boston, Massachusetts 02115}

\maketitle

\begin{abstract}
It is shown that certain sum rule identities  exist which
relate  correlation functions for $n$ Potts spins  on the boundary of
a planar lattice  for $n\geq 4$.
  Explicit expressions of the identities are obtained for $n=4,5$.
It is also shown  that the
 identities provide the missing link needed for
  a complete determination of  the duality
relation for the  $n$-point correlation function.
 The $n=4$ duality relation is obtained explicitly. More generally
 we deduce the
 number of correlation identities for any  $n$ as well as an 
inversion relation and a conjecture on the general form of
the duality relation.

\end{abstract}

\vskip 1cm
\pacs{05.50.+q}
 
The Potts model \cite{potts}, which is a generalization of the two-component 
Ising model to $q$ components for arbitrary $q$, has been a subject matter
of intense interest in many fields ranging from condensed matter
to high-energy physics.  For reviews on 
the Potts model and its relevance see, for example, \cite{wuPotts,wuapplied}.
However, exact results on the Potts model have proven to be extremely elusive.
Rigorous results known to this date are limited, and include 
essentially  only a   closed-form evaluation of its free energy 
 for  $q=2$, the Ising model \cite{onsager}, and 
critical properties 
for the
square, triangular and honeycomb lattices  \cite{baxter,bta,kunz}.
Much less    is known about its  correlation functions.

In this Letter we report on new sum rule identities for
the Potts  $n$-point correlation function.
  Specifically, we show that,
as a consequence of being a many-component system, the correlation functions
of Potts spins on the boundary of a planar lattice
must necessarily satisfy certain  identities
when $n\geq 4$. We further show that  these identities 
lead to the complete
determination of a correlation duality relation  which, in its simplest form,
has proven to be useful in determining   the equilibrium crystal shape 
of the Ising model \cite{het,zia1}.
Our results are very general and 
hold for any planar lattice or graph with arbitrary interactions. 

 Consider the $q$-state Potts model 
on a planar lattice, or graph,  ${\cal L}$ of $N$ sites and
$E$ edges. Let $i,j,\cdots,m,\ell$ be
$n$ sites on the boundary ordered as 
shown in Fig. 1, and let $\s_i$ denote the state of the spin at site $i$.
Two spins of \L
at sites $i'$ and $j'$ interact with an interaction
$K_{ij}\d(\sigma_{i'}, \sigma_{j'})$, where $\sigma_i',\s_j'=1,2,...,q$.
Define the $n$-point correlation function \cite{wu}
\be
P_n(\s, \s', \cdots, \s^{(n)}) 
=<\d(\s_i,\s)\d(\s_j,\s')\cdots\d(\s_\ell,\s^{(n)})> \label{npoint}
\ee
as the probability that the 
$n$ spins are in respective
  {\it definite} spin states
$\sigma, \s',\cdots\s^{(n)}$.
In particular, the correlation function
\be
\Gamma_n= q^n P_n(\s,\s,\cdots,\s) -1 \label{ngamma}
\ee
vanishes identically if the $n$ spins are
completely uncorrelated.  

It is convenient to write
$P_{ij \cdots \ell}= P_n(i, j, \cdots \ell) =Z_{ij \cdots \ell}/Z$, 
where $i,j,\cdots,\ell = 1,2,\cdots,q$, $Z$ is the partition function, and 
  $Z_{ij\cdots \ell}$  the partial partition function, namely, the 
sum of Boltzmann factors
with the  boundary spin states 
fixed at $i,j,\cdots,\ell$.
 Then we have the following theorem.

\noindent
{\it Theorem}:
(i) The correlation functions $P_n$, $n\geq 4$,  are related by  certain 
sum rule identities.
Particularly,
for $n=4$ and 5, the identities are 
\bea
P_{1212} & =& P_{1213} +P_{2131}-P_{1234} \label{P4}\\
P_{12112} & =& P_{12113} +P_{21331} - P_{12334}
\nonumber \\
P_{12123} & =& P_{12134} +P_{21314}- P_{12345}
 \label{P5}
\eea
and eight other relations obtained by
cyclically permuting the five indices in (\ref{P5}).
 
(ii) The number  of  correlation identities
for a given $n$ is 
$a_n = b_n-c_n $,
where  $b_n$ and $c_n$ are generated respectively from the
generating  functions
 \bea
{\rm exp} (e^t -1) &=&\sum_{n=0}^\infty b_nt^n/n! \label{gen1} \\
(1-\sqrt{1-4t})/2t&=& \sum_{n=0}^\infty c_nt^n. \label{gen2} 
\eea

\noindent
{\it Proof}:
The identity (\ref{P4}) is equivalent to
\be
Z_{1212}  =  Z_{1213}+Z_{2131}- Z_{1234}, \label{P44}
\ee
which we represent  graphically in Fig. 2.
Consider the high-temperature expansion  of $Z_{ijk\ell}$
in the form \cite{baxterwu} of
\be
Z_{ijk\ell} = 
 \sum_{G} q^{n(G)}\prod _{i',j'} \biggl(e^{K_{i'j'}}-1\biggr). \label{graph}
\ee
Here, as a consequence of the fact that the four boundary sites are fixed
in definite spin states, the  summation 
 is taken over all graphs $G\subseteq {\cal L}$ in which
 there are $n(G)$ clusters 
   excluding those connected to the four boundary
 sites.

Apply the expansion (\ref{graph}) to the four $Z$'s in (\ref{P44}).
 It is clear that, as a consequence of \L being planar,
 we have $Z_{1212}=T_1+T_2+T_3$ where $T_1$ is the sum of
graphs where sites $i$ and $k$  belong to  the same cluster,
$T_2$ those graphs where sites $j$ and $\ell$ belong
 to the same cluster, and $T_3$ graphs 
$i,j,k,\ell$ all belong to  different clusters.
It is also clear that we have $Z_{1213}= T_1+T_3, \>Z_{2131}=T_2+T_3$
and $Z_{1234}=T_3$.
 The identity (\ref{P44}), and
hence (\ref{P4}), now follows as a sum rule condition.
Clearly, the existence of (\ref{P4}) is a consequence of
the planar connectivity topology.    
It can also be  shown that all $n=5$ identities 
are generated  by inserting
one boundary site to the 
diagrams in Fig. 2, resulting in identities (\ref{P5}) shown graphically 
in Fig. 3.
One can proceed in a similar fashion to derive sum rules for
$n\geq 6$, and thus we have
established (i).
We remark that the sum rules manifest themselves 
 only
for $q\geq 4$, and therefore do not apply to  the Ising model.

 To
 enumerate  $a_n$, the number of 
correlation identities for a given $n$,
it is instructive to consider 
the case $n=4$.
First, by enumeration we find that
 there are 15 distinct $Z_{ijk\ell}$.  For
each $Z_{ijk\ell}$ we construct its graph as in Fig. 2 
and connect sites in the same state by drawing
connecting lines exterior to ${\cal L}$, resulting in
a ``connectivity" of the four points.
(There is no distinction in connectivity topology
between drawing connecting lines  within or exterior to ${\cal L}$).
A well-nested connectivity, or  well-nested $Z$ for brevity, is one in 
which the 
connecting lines do not intersect \cite{bn}.
For $n=4$, 14 of the 15 Z's, which are  shown in Fig. 4,  are  well-nested.
Only $Z_{1212}$ which, for precisely the same planar  topology reason
noted in the above, is not well-nested.

More generally for a given $n$-point correlation function
$Z_{ij\cdots m\ell}$, or $Z$ for brevity, one  connects 
in its graph sites in the same state to arrive
at an $n$-point connectivity.
Let there be altogether $b_n$ distinct connectivities of which
$c_n$ are well-nested.   To each $Z$ which 
 is not well-nested, we  follow the procedure describe
in the above, namely, expanding graphically in a high-temperature
series.  Since all graphs in the expansion
do not contain intersecting lines, by applying the
principle of inclusion-exclusion \cite{birkhoff} we 
eventually arrive at a sum rule expressing  the particular
correlation function in question in terms of
well-nested ones.  This gives rise to an
identity for this particular $Z$.
Furthermore, since each $Z$ has a unique graphical expansion, 
all identities are distinct.  It follows that 
the number of sum rule identities, $a_n$,
is equal to the number of $Z$'s
which are not well-nested, namely, $b_n-c_n$.
 
The number $c_n$
 has been evaluated by Bl\"ote and Nightingale \cite{bn}
in a consideration of the transfer matrix formulation of the Potts model,
and is found to be generated by (\ref{gen2}). 
  To enumerate  $b_n$ we note that  it is  
precisely the number of ways that $n$ objects can be
partitioned into  indistinguishable 
parts.
 Let there be $m_\nu$ parts of $\nu$ objects each, $\nu = 1,2,\cdots$.
Then we have
$b_n = \sum_{m_\nu =0}^\infty\ '  \prod_{\nu=1}^\infty
[(n!)/{(\nu!)^{m_\nu}m_\mu !  }] $
where the prime over the summation indicates
the condition $\sum_{\nu=1}^\infty \nu m_\nu =n$.
This leads to 
the generating function (\ref{gen1})
and thus establishes (ii).  
Particularly,  we find
$a_4=15-14=1, a_5=52-42=10, a_6=203-132 =71 , a_7= 877-429=448.$
We have verified these numbers by explicitly enumerating all  
connectivities for $n\leq 6$.

\medskip
\noindent
{\it Duality relation for $P_n$}:

\noindent
 
It  has been known for some time that the two-point correlation function of
an  Ising model  is related to
 its counterpart in 
  the dual space.
 The usual derivation of this relation involves
embedding expansions of the correlation
functions
 on the lattice followed by an  explicit term-by-term identification
\cite{watson,zia}.  In a recent paper
one of us \cite{wu} introduced a new approach 
to this problem which invokes only
a repeated use of an elementary
duality consideration \cite{wuwang}.
The new approach, which is very general,  also permits the extension of 
 the duality analysis to the Potts model for $n=2,3$ \cite{wu}.
However, an extension of the analysis of \cite{wu} to $n\geq 4$
ran into an   apparent snag of  inadequacy of 
conditions \cite{jacobsen}.
  Here we show  that the correlation identities 
derived in the above provide the missing link,
and with the help of these identities we
 determine the 
 duality relation for any $n$.

The consideration of \cite{wu} is based on 
the fundamental duality relation
 \cite{wuwang} 
\be
Z=qCZ^* \label{part}
\ee
relating the partition function $Z$ of any planar lattice, or graph, 
 to the partition function $Z^*$ 
 on the dual.
  Here, 
   $C= q^{-N^*}\prod_{\rm edges} (e^{K_{ij}}-1)$, with 
 $N^*$ being the number of sites of the dual
and the product taken over all edges. 
The interaction $K^*_{ij}$  dual to $K_{ij}$
is given by
\be
(e^{K_{ij}}-1)(e^{K^*_{ij}}-1) = q. \label{dual}
\ee

Starting from \L we
 consider a lattice ${\cal L}^*$ 
formed by introducing
$n$  spins $\a,\b,\c\cdots,\d$ to the boundary of the dual of \L
(Cf. Fig. 1), each interacting with neighboring dual spins within
${\cal L}$. (Note that ${\cal L}^*$ now has $N^*+n-1$ sites and
is not the dual of ${\cal L}$.)
 Let $Z^*_{\a\b\c\cdots\d}$ be
the partial dual partition function of ${\cal L}^*$
with the $n$ boundary spins fixed in
the respective definite states.  Our goal is to obtain 
a duality relation in the form of a linear transformation relating the 
$Z_{ij\cdots m\ell}$ to $Z^*_{\a\b\c \cdots\d}$.

Regard the $b_n$ well-nested connectivities (such as those
shown in Fig. 4 for $n=4$) as
auxiliary lattices, and apply the fundamental duality relation to 
each one of them \cite{jacobsen}.
Applying the duality on ${\cal L}$ itself, for example, we obtain
(\ref{part}) which can be written as
an equation relating  linear combinations of the $Z$ and $Z^*$ \cite{wu}.
Applying the duality to the well-nested connectivity ${\cal L}_n$
in which all $n$ points are connected to a common point
with interactions $K$  as in  ${\cal L}_4$  shown in Fig. 4(b), 
 we obtain
  \be
Z_{{\rm aux}(n)}=\biggl({{qC}\over {q^{n-1}}}
\biggr) (u-1)^n \>Z^*_{{\rm aux}(n)},
\label{part1}
\ee
where $u=e^K$, and $Z_{{\rm aux}(n)}$ and
$Z^*_{{\rm aux}(n)}$ are respectively the partition functions of
${\cal L}_n$ and its dual.
Now, both sides of (\ref{part1}) are polynomials of degree
$n$ in $u$.
Since (\ref{part1}) holds for arbitrary $u$,
the coefficients of all powers of $u$ 
must be equal.  However, it suffices to equate only the coefficients of the
highest power of $u$
(equating   other coefficients
leads simply to linear combinations of equations to be obtained from other 
connectivities).
On the LHS we have 
 $Z_{{\rm aux}(n)} =q(u+q-1)^nZ_{11\cdots 1} +O(u^{n-1})$, and find the
 coefficient
of $u^n$ to be $qZ_{11\dots 1}$.  On the RHS we have $(u-1)^nZ^*_{{\rm
aux}(n)}=
q(u+q-1)^nZ^*_{11\cdots 1} +$ other terms, after using (\ref{dual}), and
 the coefficient of $u^n$ is    a linear combination of the $b_n$
$Z^*$.  This leads immediately to an expression for $Z_{11\cdots 1}$.
For $n=4$, for example, we obtain
  \bea
Z_{1111}& = &{C\over {q^{2}}}
\biggl[ Z^*_{1111}+q_1(Z^*_{2111}+Z^*_{1211}+Z^*_{1121}+Z^*_{1112}) 
 +q_1(Z^*_{1122}+Z^*_{1221}) \nonumber \\
&& 
 +q_1q_2(Z^*_{1123}+Z^*_{2113}+Z^*_{2311}+Z^*_{1231} ) 
 +q_1( Z^*_{1212}+q_2Z^*_{1213}+q_2Z^*_{2131}
 +q_2q_3Z^*_{1234})\biggr]\nonumber\\
&=& \{1+q_1(1,1,1,1)+ q_1(1,1)+ q_1q_2(1,1,1,1)
 + q_1(1,q_2,q_2,q_2q_3)\}, \label{1111}
\eea
where $q_m=q-m, m=1,2,\cdots$, and
in the last line we have introduced a short-handed notation.
An immediate consequence of (\ref{1111}) is the result
 \bea
\Gamma_4& =&q_1(p_{2111}+p_{1211}+p_{1121}+p_{1112}+p_{1212}+p_{1122}
+p_{1221})  \nonumber \\
&& +q_1q_2(p_{1123}+p_{2113}+p_{2311}+p_{1231}+p_{1213}+p_{2131})
+ q_1q_2q_3p_{1234},
\label{gamma}
\eea
where we have introduced (\ref{part}), $Z^*=qZ^*_{1111}$, as well as
$p_{\a\b\c\d} = Z^*_{\a\b\c\d}/Z^*_{1111}$.
For general $n$ the consideration of ${\cal L}_n$   leads to
\be 
\Gamma_n = q_1\sum p_{211\cdots 1} +q_1q_2\sum p_{231\cdots 1}
+\cdots +q_1q_2\cdots q_{n-1} p_{123\cdots n} \label{gn}
\ee
where  the meaning of the summations is obvious. 
  
 Applying (\ref{dual}) to all $c_n$
auxiliary lattices  of  well-nested
connectivities in this fashion
 and equating the coefficients of the highest power of
$u$ in each case,
  we obtain $c_n$ equations  for
the $b_n$ unknown $Z$'s.
 Since $c_n<b_n$ for $n\geq 4$, 
it appears that there are more unknowns than equations and that
the equations are inadequate  \cite{jacobsen}.
However, after combining the $c_n$ equations with the $b_n-c_n$ sum rule
identities,
we have precisely $b_n$ equations, and
the duality relation can now be 
completely determined!

In the case of  $n=4$,  the solution of the 15 equations leads to,
in addition to  (\ref{1111}),
  \bea
Z_{2111}& =& \{1+(-1, -1,q_1,q_1)+ (q_1,-1) +q_2(q_1,-1,-1,-1)  
  -(1,q_2,q_2,q_2q_3)\} \nonumber \\
  Z_{1122}& =& \{1+(-1,q_1,-1,q_1) -(1,1) -q_2(1,1,1,1)
+(q_1,q_1q_2,-q_2,- q_2q_3) \} \nonumber \\
  Z_{1123}& =& \{1-(1,-q_1,1,1) -(1,1)  +(2,2,-q_2,-q_2)
 +(-1,-q_2,2,2q_3) \} \nonumber \\
  Z_{1212} &=& \{1-(1,1,1,1) +q_1(1,1) -q_2(1,1,1,1)
+Q(-1,q_1q_2, q_1q_2,q_2q_3r)\} \nonumber \\
 Z_{1213} &=& \{1-(1,1,1,1) +(-1,q_1) +(2,-q_2,2,-q_2) 
+Q(q_1,s,s,q_3t)\} \nonumber \\
 Z_{2131} &=& \{1-(1,1,1,1) +(q_1,-1) +(-q_2,2,-q_2,2) 
 +Q(q_1,s,s,q_3t)\} \nonumber \\
 Z_{1234} &=& \{1-(1,1,1,1) -(1,1) +2(1,1,1,1) 
 +Q[r,t,t,q_3(2-5q)]\} \label{all}
\eea
where $Q=1/(q^2-3q+1), r=2q-1, s=q^2-4q+2, t=q^2-5q+2$.
Expressions for $\{ Z_{1211}, Z_{1121}, Z_{1112} \}, \{Z_{1221}\}$
and $\{Z_{2113}, Z_{2311} , Z_{1231}\}$ are given by cyclic
permutations.
We remark that a closer examination shows that all $Z_{ij k\ell}$ 
except the last
four in (\ref{all}) can be determined without the use of the identity
(\ref{P44}).   

The solution (\ref{1111}) and (\ref{all})
 can be written more compactly by using the fact that
$Z^*$, as partial partition functions of ${\cal L}^*$,
satisfy the same sum rules as the $Z$. Particularly, 
we have
$Z^*_{1212} = Z^*_{1213} +Z^*_{2131}
-Z^*_{1234}$.  
 Using this relation and rewriting (\ref{npoint}) as
 \bea
&&P_4(\s_1, \s_2, \s_3, \s_4)= A_{1234} 
+ A_{1123}\d_{12}
+A_{2113}\d_{23} +A_{2311}\d_{34} +A_{1231} \d_{14} \nonumber \\
&&\hskip .7cm +A_{1213}\d_{13}
+A_{2131}\d_{24}
+A_{1122}\d_{12}\d_{34} +A_{1221}\d_{14}\d_{23} 
+A_{1212}\d_{13}\d_{24}\nonumber\\
&& \hskip .7cm
+A_{2111}\d_{234}+A_{1211}\d_{134}+A_{1121}\d_{124}+A_{1112}\d_{123}
+A_{1111}\d_{1234}
 \label{4spin}
\eea
where $\d_{12}=\d(\s_1, \s_2), \d_{ 123} =\d_{12}\d_{23}$, etc.
we obtain after some algebra
  \begin{eqnarray}
A_{1234}&=&q^{-4}[1-
(p_{2111}+p_{1211}+p_{1121}+p_{1112}+p_{1122}+p_{1221}+p_{1212})\nonumber \\
        &&\hskip 1.3cm  +2(p_{1123}+p_{1231}+p_{2113}
+p_{2311}+p_{1213}+p_{2131}) 
      -6p_{1234}] \nonumber \\ 
A_{1123}&=&q^{-3}(p_{1211}-p_{1231}-p_{2311}-p_{1213}+2p_{1234}) \nonumber \\
A_{2113}&=&q^{-3}(p_{1121}-p_{1123}-p_{1231}-p_{2131}+2p_{1234}) \nonumber \\
A_{2311}&=&q^{-3}(p_{1112}-p_{2113}-p_{1123}-p_{1213}+2p_{1234}) \nonumber \\
A_{1231}&=&q^{-3}(p_{2111}-p_{2311}-p_{2113}-p_{2131}+2p_{1234}) \nonumber \\
A_{1213}&=&q^{-3}(p_{1221}-p_{1231}-p_{2113}+p_{1234}), \hskip .5cm
A_{2131}=q^{-3}(p_{1122}-p_{1123}-p_{2311}+p_{1234})\nonumber \\ 
A_{1122}&=&q^{-2} (p_{1213}-p_{1234}),\hskip .5cm
A_{1221}=q^{-2}(p_{2131}-p_{1234}),\hskip .5cm 
A_{1212}=0\nonumber \\
A_{2111}&=&q^{-2} (p_{1123}-p_{1234}),\hskip .5cm 
A_{1211}=q^{-2} (p_{2113}-p_{1234})\nonumber \\
A_{1121}&=&q^{-2} (p_{2311}-p_{1234}),\hskip .5cm
A_{1112}=q^{-2} (p_{1231}-p_{1234}),
\hskip .5cm A_{1111}=q^{-1} p_{1234}. \label{as}
\end{eqnarray}
For general $n$ we write in analogous to 
(\ref{4spin})
\be 
P_n(\s_1,\s_2,\cdots,\s_n) = A_{12\cdots n} +A_{1123\cdots(n-1)}\d_{12}
+\cdots +A_{11\cdots 1} \d_{12\cdots n} \label{aa}
\ee
and similarly for boundary spins $\s_\a, \s_\b,\cdots, \s_\d$ of
${\cal L}^*$.
\be 
P^*_n(\s_\a,\s_\b,\cdots,\s_\d) = A^*_{12\cdots n}
+A^*_{1123\cdots(n-1)}\d_{\a\b}
+\cdots +A^*_{11\cdots 1} \d_{\a\b\cdots \d}.\label{aaa}
\ee
Regard the diagram in Fig. 1 as representing $A_{ij\cdots \ell}$ and
construct for each $A$ the associated connectivity as previously
described.  Then we are led to the following working conjecture:

\noindent
{\it Conjecture}:
\bea
A_{ij\cdots \ell} &=& q^{1-m} A^*_{\a\b\cdots \d} \hskip 1.5cm
{\rm if \>\>the \>\>connecitivity\>\>is\>\>well}-{\rm nested}, \nonumber\\
&=& 0, \hskip 3cm {\rm otherwise},
\eea
where $m$ is the number of distinct indices in $\{i,j,\cdots,\ell\}$.
The conjecture is readily verified  for $n=2,3,4$.
In practice, for any given $n$, one can 
solve  $ A^*_{\a\b\cdots \d} $ from (\ref{aaa}) in terms 
of $ P^*_{\a\b\cdots \d} =q^{-1}p^*_{\a\b\cdots\d}$
by applying the principle of inclusion-exclusion \cite{birkhoff}.
Details will not be given.

\smallskip
\noindent
{\it An inversion relation}: 
Since ${\a\b\c\cdots\d}$ are boundary sites of 
${\cal L}^*$,
 the  transformation relating $Z^*$ 
to $Z$, an inversion process, is given precisely by the same transformation
relating $Z$ to $Z^*$.
Now ${\cal L}^*$ has $N^*+n-1$ sites and its dual
has $N-n+1$ sites. Also  \L and ${\cal L}^*$ have the same number of
edges.  Therefore, we have
\bea
Z_{ijk\cdots m\ell} &=& \biggl({{\prod(e^{K_{ij}}-1)}\over
{q^{N^*+(n-2)}}}\biggr)
\sum_{\a\b\c\cdots\d} {\bf M}(ij\cdots m\ell|\a\b\c\cdots\d)
Z^*_{\a\b\c\cdots\d}, \label{matrix}
\\
Z^*_{\a\b\c\cdots\d}& =& \biggl({{\prod(e^{K^*_{ij}}-1)}\over
{q^{N-n+1+(n-2)}}}\biggr)
\sum_{ij\cdots m\ell} {\bf M}(\a\b\c\cdots\d|\ell ij\cdots m) 
Z_{\ell ij\dots m}, \label{invmatrix}
\eea
where ${\bf M}$ is a $b_n\times b_n$ matrix.
From Fig. 1 we observe that the resulting spin indices 
on \L after the inversion
is a counter-clockwise cyclic permutation of the original ordering.
Substituting (\ref{invmatrix}) into (\ref{matrix})
and making use of (\ref{dual}) and the Euler relation
$E=N + N^* -2,$
we are  led to the identity 
\be 
{\bf M}^2 (ijk\cdots\ell|i'j'k'\cdots\ell') = q^{n-1} \d_{ij'}\d_{jk'}
\cdots\d_{\ell i'}, \label{i}
\ee
which we refer to as an inversion relation.
We have explicitly verified 
   this  inversion relation  for $n=2,3,4$.
It can also be shown that the $n=4$ inversion relation can be used to
deduce the sum rule identity (\ref{P44}).

We are grateful to J. L. Jacobsen  for  discussions
and a comment \cite{jacobsen}
   which has led to this investigation.
 This work is supported in part by the National Science Foundation
Grant DMR-9614170.

\begin{center}

{\bf Figure captions}

\end{center}

\noindent
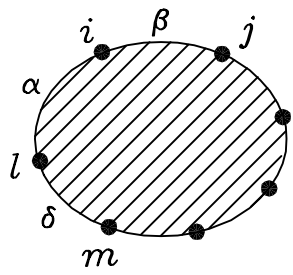
Fig. 1.
 A planar lattice \L and $n$ sites $i,j,\dots m,\ell$ on the boundary.

\medskip
\noindent
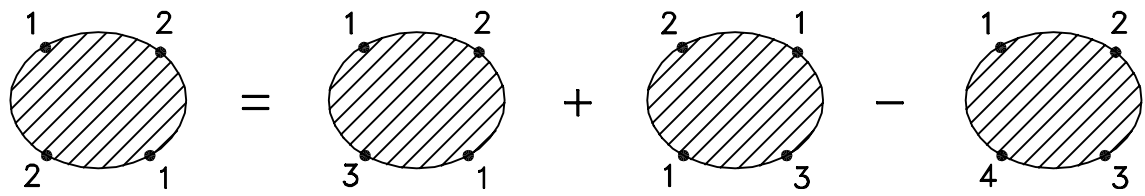
Fig. 2. Graphical representation of the sum rule identity (\ref{P4}).

\medskip
\noindent
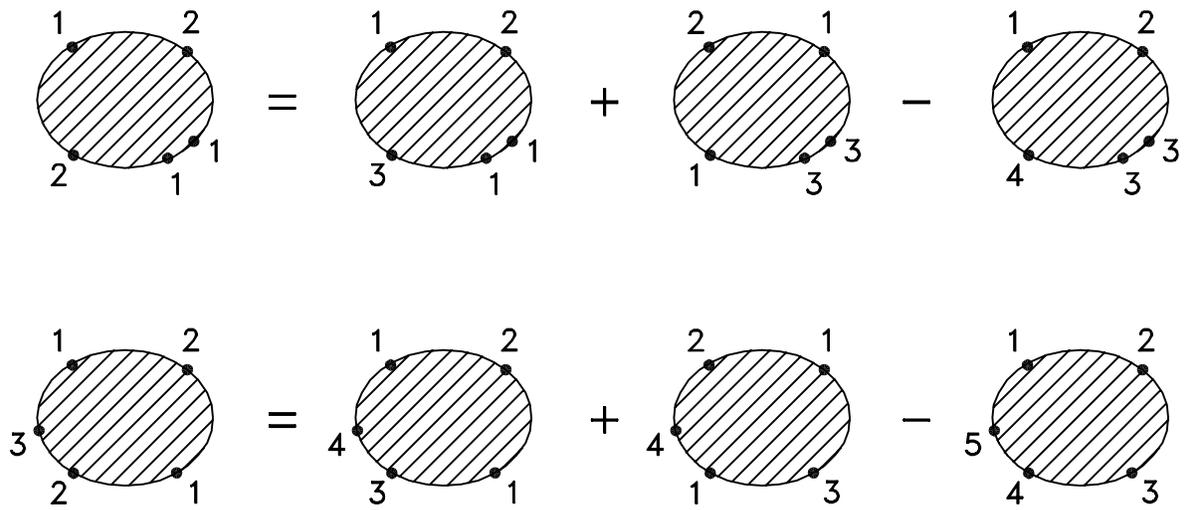
Fig. 3. Graphical representation of the sum rule identities
(\ref{P5}).

\medskip
\noindent
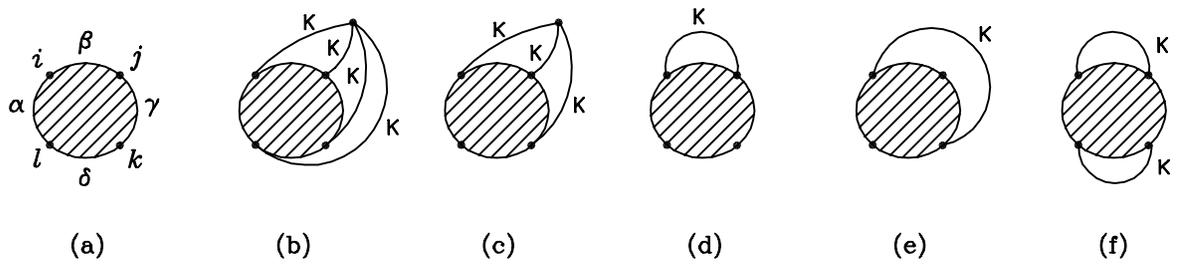
Fig. 4. The  14 well-nested connectivities for $n=4$
corresponding to (a) $Z_{1234}$, (b) $Z_{1111}$, (c) 
$Z_{1112}$ occurring  4 times, (d) $Z_{1123}$  occurring  4 times,
(e)  $Z_{1213}$ occurring  2 times, and (f) $Z_{1122}$ occurring  2 times.

\end{document}